\documentclass[fleqn,12pt,twoside]{article}
\usepackage[headings]{espcrc1}

\readRCS
$Id: espcrc1.tex,v 1.2 2004/02/24 11:22:11 spepping Exp $
\usepackage{graphicx}
\usepackage[figuresright]{rotating}

\newcommand{\ba}{\begin{eqnarray}}
\newcommand{\ea}{\end{eqnarray}}
\newcommand{\bmath}{\begin{mathletters}}
\newcommand{\emath}{\end{mathletters}}
\newcommand{\ban}{\begin{eqnarray*}}
\newcommand{\ean}{\end{eqnarray*}}

\newcommand{\AmS}{{\protect\the\textfont2
  A\kern-.1667em\lower.5ex\hbox{M}\kern-.125emS}}

\title{Quantum Shape-Phase Transitions in Finite Nuclei}

\author{A. Leviatan
        \address{Racah Institute of Physics, 
        The Hebrew University, Jerusalem 91904, Israel}%
        \thanks{This work is supported by the Israel Science Foundation.}}
\runtitle{Quantum shape-phase transitions in finite nuclei}
\runauthor{A. Leviatan}

\begin{document}

\maketitle

\begin{abstract}
Quantum shape-phase transitions in finite nuclei are considered in the 
framework of the interacting boson model. Critical-point Hamiltonians 
for first- and second-order transitions are identified by resolving them 
into intrinsic and collective parts. 
Suitable wave functions and finite-N estimates for 
observables at the critical-points are derived. 
\end{abstract}

\section{QUANTUM PHASE TRANSITIONS}

Quantum phase transitions are structural changes occurring at
zero temperature as a function of a coupling constant 
and driven by quantum fluctuations. 
An example is the phase transition between different shapes of 
nuclei for which the equilibrium deformation plays the role of an 
order parameter. 
A proper description of such phase transitions 
necessitates the inclusion of correlation effects beyond the 
mean-field level. In a microscopic approach this is done by 
restoring broken symmetries of the self-consistent mean-field solutions 
(HFB or HF+BCS) 
via angular momentum and number projection, and by 
configuration mixing using the generator coordinate method~\cite{bender03}.
Shape-phase transitions in nuclei have also been studied 
in the geometric framework of a Bohr Hamiltonian for macroscopic 
quadrupole shapes. Analytic solutions, 
called E(5)~\cite{iac00} and X(5)~\cite{iac01}, were obtained under certain 
approximations ({\it e.g.} infinite square-well potentials), and 
shown to be relevant to 
nuclei at the critical-points of second- and (low-barrier) 
first-order transitions respectively~\cite{caszam0001}. 
A key issue 
near criticality is to understand the modifications brought 
in by the fact that nuclei consist of a finite number of nucleons. 
This aspect is particularly important for sorting out the 
phase diagram and structure-evolution of stable and unstable nuclei, 
as a function of protons and neutrons. 
In the present contribution we 
examine this question  
in the framework of the interacting 
boson model (IBM)~[5] which describes low-lying quadrupole collective 
states in nuclei in terms of a system of $N$ monopole ($s$) and
quadrupole ($d$) bosons representing valence nucleon pairs. 
This algebraic model 
is simple yet rich enough to 
encompass the dynamics of 
quantum shape-phase transitions in nuclei, and 
illuminate the underlying finite-N structure 
at the critical-point. 

\section{GEOMETRY AND CRITICAL-POINT HAMILTONIANS}

The starting point for analyzing shapes in the IBM is 
an energy surface defined by 
the expectation value of the Hamiltonian in the coherent (intrinsic) 
state~\cite{gino80,diep80}
\ba
\vert\,\beta,\gamma ; N \rangle &=&
(N!)^{-1/2}(b^{\dagger}_{c})^N\,\vert 0\,\rangle ~,
\label{cond}
\ea
where $b^{\dagger}_{c} = (1+\beta^2)^{-1/2}[\beta\cos\gamma 
d^{\dagger}_{0} + \beta\sin{\gamma} 
( d^{\dagger}_{2} + d^{\dagger}_{-2})/\sqrt{2} + s^{\dagger}]$.
For the general IBM Hamiltonian with one- and two-body interactions, the 
energy surface takes the form
\ba
E_{N}(\beta,\gamma) = E_0 + 
N(N-1)
\left [ a\beta^{2} - b\beta^3\cos 3\gamma + c\beta^4\right ]
(1+\beta^2)^{-2} ~.
\,
\label{eint}
\ea
The coefficients $E_0,a,b,c$ involve particular linear 
combinations of the Hamiltonian's parameters~\cite{kirlev85,lev87}. 
The equilibrium shape for a given Hamiltonian is determined by the global 
minimum, $(\beta_{eq},\gamma_{eq})$, of the energy surface, and can be 
spherical ($\beta_{eq}=0$) or deformed $(\beta_{eq}>0$) with 
prolate ($b>0$), oblate ($b<0$) and $\gamma$-independent ($b=0$) character.
For $\beta>0$ the intrinsic state of Eq.~(\ref{cond}) is deformed, 
and represents a band  
whose rotational members are obtained by projection. For
example, for $\gamma=0$ the band consists of states with good $O(3)$ 
symmetry $L=0,2,4,\ldots,2N$, given by
\ba
\vert\, \beta; N, L,M\rangle &\propto&
\hat{\cal{P}}_{LM}\vert \beta,\gamma=0; N\rangle ~.
\label{wfqpt1}
\ea
The projected states with fixed $N$ and $L$, involve a mixture 
of components $\vert N,n_d,\tau,L\rangle$ with quantum numbers related 
to the $U(6)\supset U(5)\supset O(5)\supset O(3)$ chain.
For $\beta=0$ one recovers 
the $U(5)$ spherical ground state, 
$\vert s^N\rangle\equiv \vert N,n_d=\tau=L=0\rangle$. 
When the deformed shape is $\gamma$-independent, 
a projection on $O(5)$ symmetry $\tau$ is required
\ba
\vert\, \beta; N,\tau, L,M\rangle &\propto& 
\hat{\cal{P}}_{\tau,LM}\vert \beta,\gamma; N\rangle ~.
\label{wfqpt2}
\ea

Stable shapes are characterized by a deep well-localized 
single minimum in the energy surface. 
In this case, the projection after variation 
procedure mentioned above 
provides an accurate approximation to 
members of the ground band. 
The dynamical symmetry limits of the IBM
are examples of such stable structures, corresponding to a 
$U(5)$ spherical vibrator ($\beta_{eq}=0$), $SU(3)$ deformed-rotor 
[($\beta_{eq}=\sqrt{2},\gamma_{eq}=0)$] and an $O(6)$ $\gamma$-independent 
rotor ($\beta_{eq}=1$). 
At a phase transition the 
energy surface serves as a Landau's potential and displays 
a different topology. 
In a first-order phase transition it 
has two coexisting minima which become degenerate at the 
critical-point. In a second-order phase transition the energy surface 
changes continuously from one minimum to another, becoming 
flat-bottomed at the critical-point. Typical examples are shown in Fig.~1. 
The corresponding conditions on 
the energy surface at the critical-points are 
\ba
&&1^{st}\, {\rm order}\qquad\;\;\;
b^{2}=4ac,\;a>0,\; b\neq 0 
\nonumber\\
&&2^{nd}\, {\rm order}\qquad\;\;\;
\, a=0,\; b=0,\; c>0 ~.
\label{1st2nd}
\ea

Phase transitions for finite N can be studied 
in the IBM by an Hamiltonian of the form, $H_1 + gH_2$, 
involving terms from different dynamical symmetry chains~\cite{diep80}. 
The critical value of the control parameter $g_c$ is determined by the 
corresponding condition in Eq.~(\ref{1st2nd}). Several studies of this 
type have identified a $U(5)$-$SU(3)$ first-order transition with 
an extremely low-barrier, a $U(5)$-$O(6)$ second-order transition, and 
a $O(6)$-$SU(3)$ cross-over~\cite{diep80,iaczam04}.
To consider shape-phase transitions of a general character, 
without being restricted to a particular 
form for the Hamiltonian, it is 
convenient to resolve the critical-Hamiltonian into intrinsic and 
collective parts~\cite{kirlev85,lev87},
\ba
H_{cri} = H_{int} + H_c ~.
\label{resol}
\ea
\begin{figure}[t]
\begin{center}
{\includegraphics[angle=270,scale=0.4]{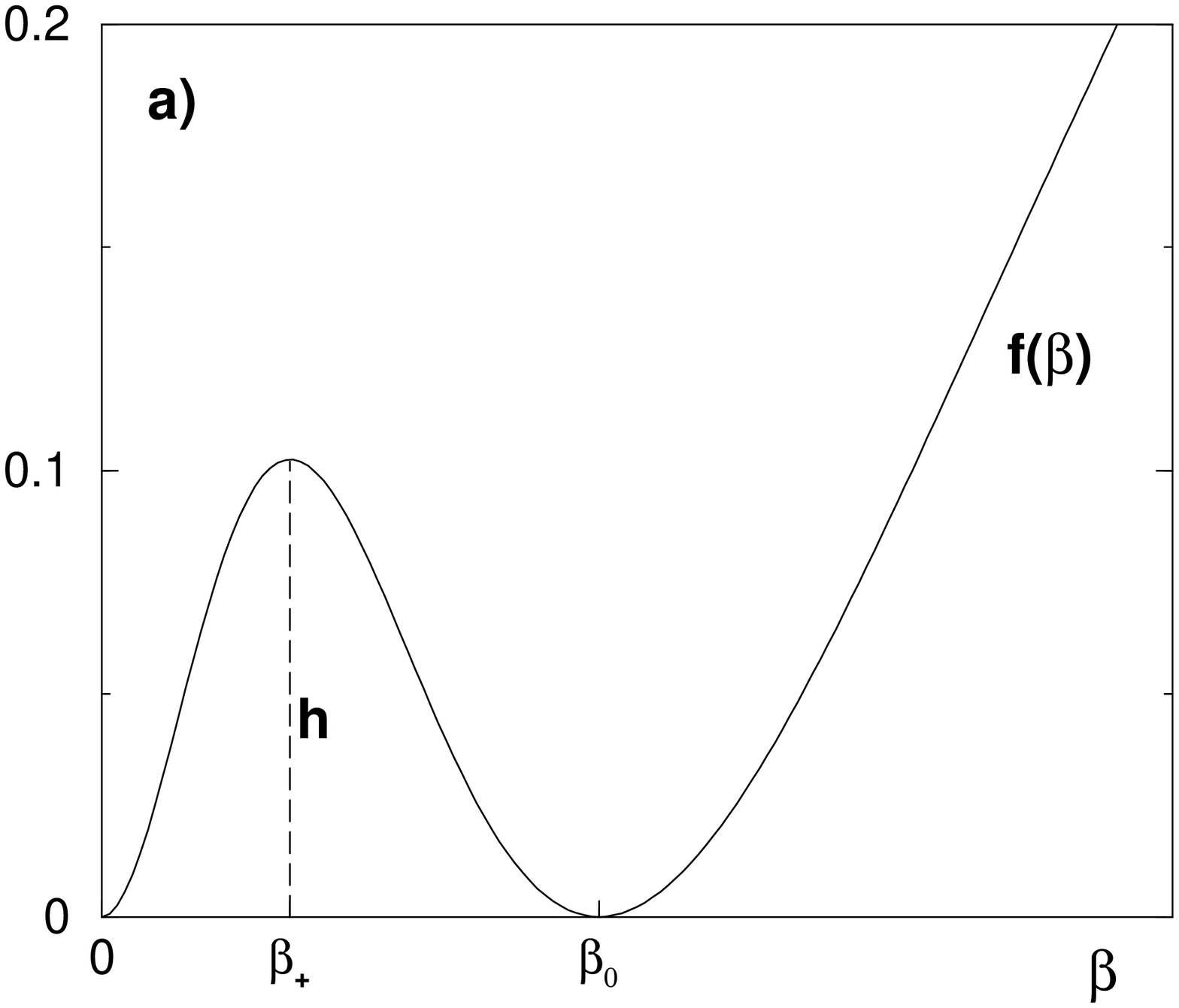}}\hspace{0.4cm}
{\includegraphics[angle=270,scale=0.4]{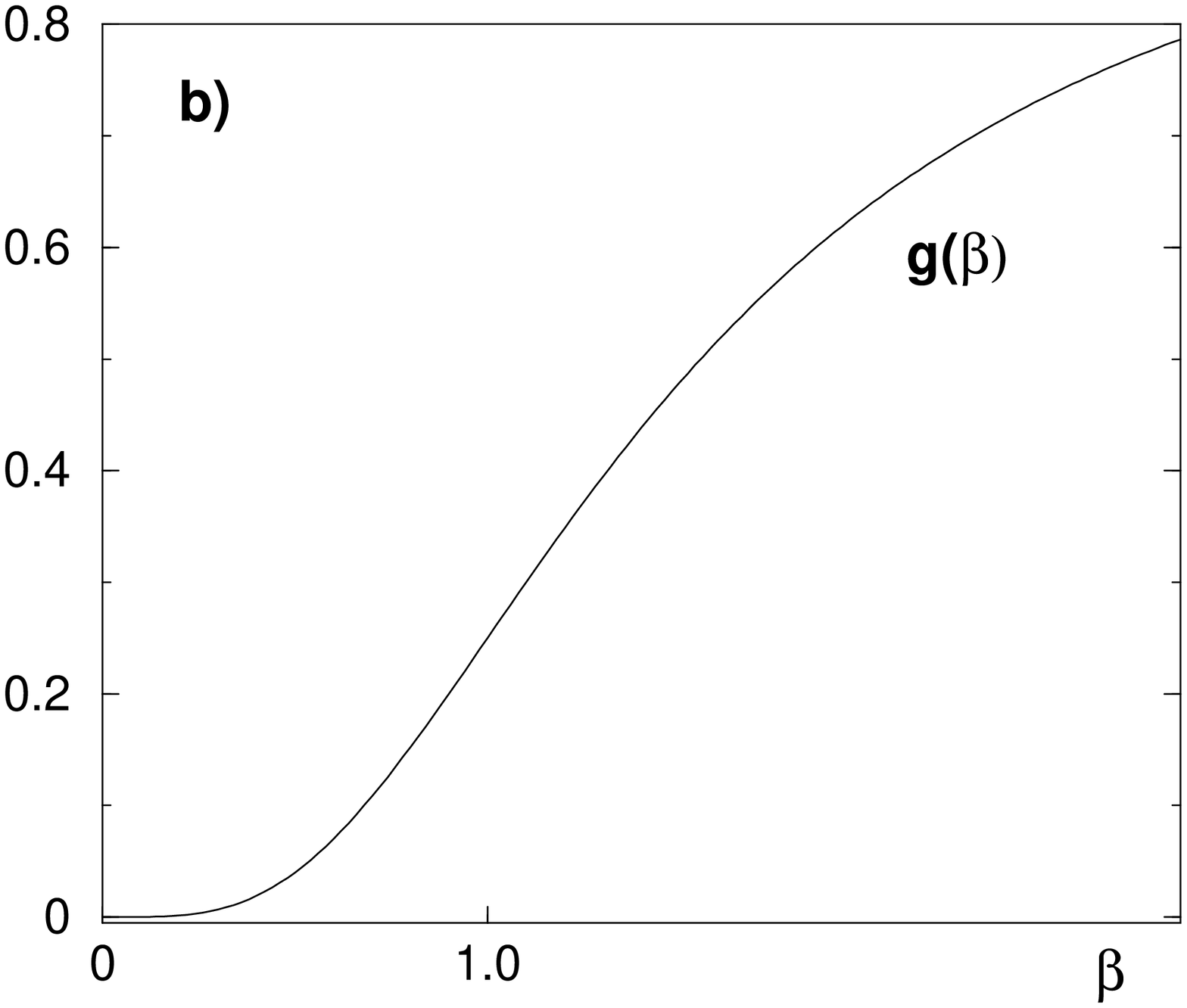}}
\end{center}
\vspace{-1cm}
\caption{Energy surfaces at the critical-points.
(a)~First-order transition, 
Eq.~(\ref{ecri1st}). (b)~Second-order transition, Eq.~(\ref{ecri2nd}). 
Asymptotically, $f(\infty)=g(\infty)=1.$}
\label{fig:1}
\end{figure}
The intrinsic part 
($H_{int}$) 
is defined to have the equilibrium 
condensate $\vert \,\beta=\beta_{eq},\gamma_{eq} ; N\rangle$, 
Eq.~(\ref{cond}), as an exact zero-energy eigenstate and to have an 
energy surface with the same shape  as the critical energy surface. 
The collective part ($H_c$) 
is composed of kinetic terms which do not affect the shape of the 
energy surface, and can be transcribed in the form
\ba
H_{c} &=& c_3 \left [\, \hat{C}_{O(3)} - 6\hat{n}_d \,\right ]
+ c_5 \left [\, \hat{C}_{O(5)} - 4\hat{n}_d \,\right ]
+\, c_6 \left [\, \hat{C}_{\overline{O(6)}} - 5\hat{N}\,\right ] +E_0~.
\label{hcol}
\ea
Here $\hat{N}=\hat{n}_d+\hat{n}_s$, $\hat{n}_d$ 
and $\hat{n}_s$ are
the total-boson, $d$-boson and $s$-boson 
number operators respectively.
$\hat{C}_{G}$ denotes the quadratic Casimir operator of the 
group G as defined in~\cite{lev87}. 
The intrinsic-collective resolution 
constitutes an efficient method for 
studying shape-phase transitions, since the derived 
Hamiltonian is tailored to reproduce a given energy surface 
which, in-turn, governs the nature of the phase transition. 

\section{FIRST-ORDER CRITICAL-POINT}
 
The energy surface at the critical-point of a 
first-order transition between spherical and prolate-deformed shapes 
has the following form for $\gamma=0$
\ba
E_{cri}(\beta) &=&
E_0 + c\,N(N-1)f(\beta)\;\;\; , \;\;\;
f(\beta) = \beta^2\,(1+\beta^2)^{-2}\,(\beta - \beta_0)^2 ~.
\label{ecri1st}
\ea
As shown in Fig.~1(a), $E_{cri}(\beta)$ has two degenerate minima
at $\beta=0$ and $\beta=\beta_0 = 2a/b = b/2c >0$. 
The value of $\beta_0$ controls the position, 
$\beta= \beta_{+} = (-1 + \sqrt{1+\beta_{0}^2}\;)/\beta_0$, 
and height $h = f(\beta_{+}) = 
( -1 + \sqrt{1+\beta_{0}^2}\,)^2/4$ of the barrier. 
\begin{figure}[t]
\begin{center}
{\includegraphics[angle=270,scale=0.38]{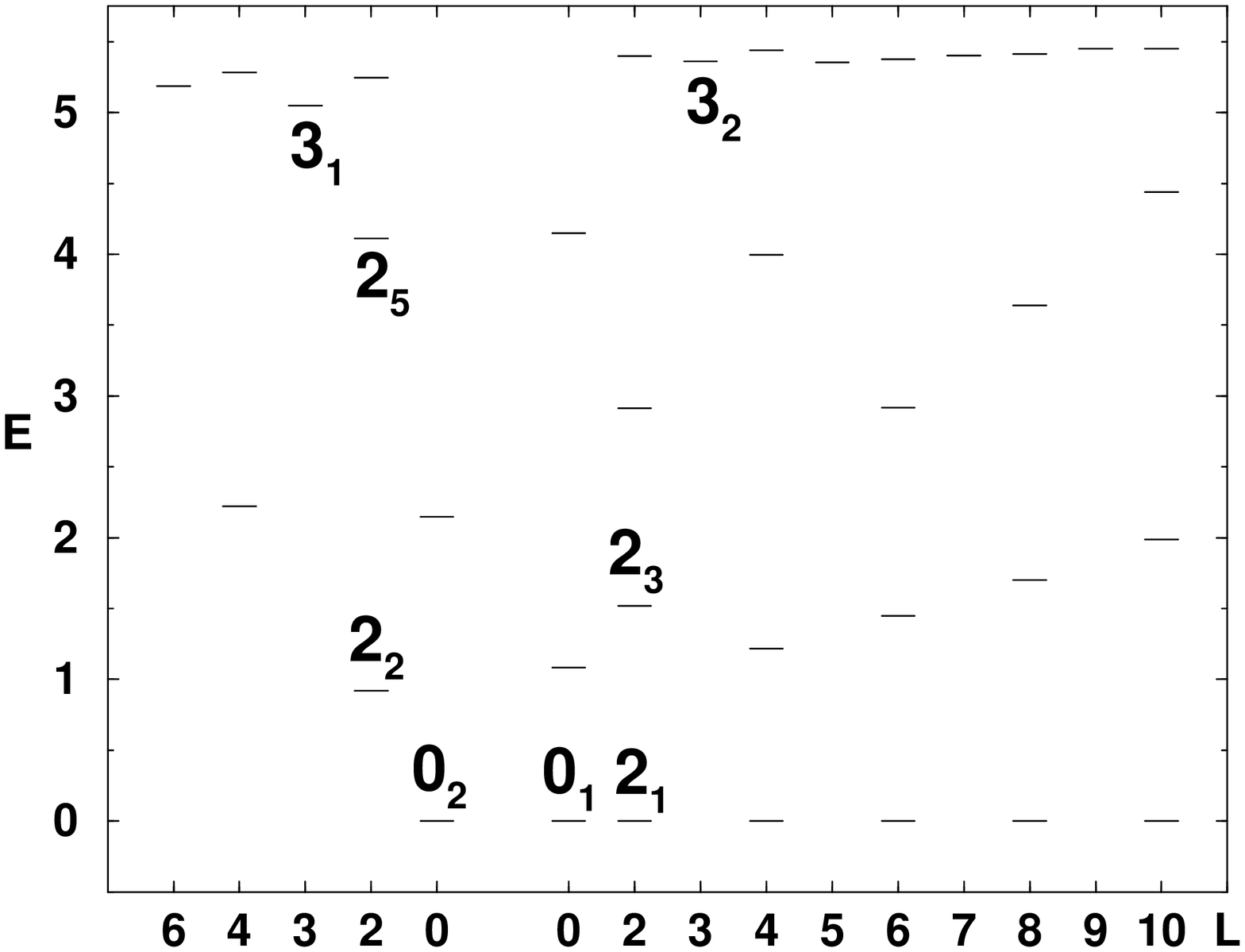}}\hspace{0.4cm}
{\includegraphics[angle=270,scale=0.38]{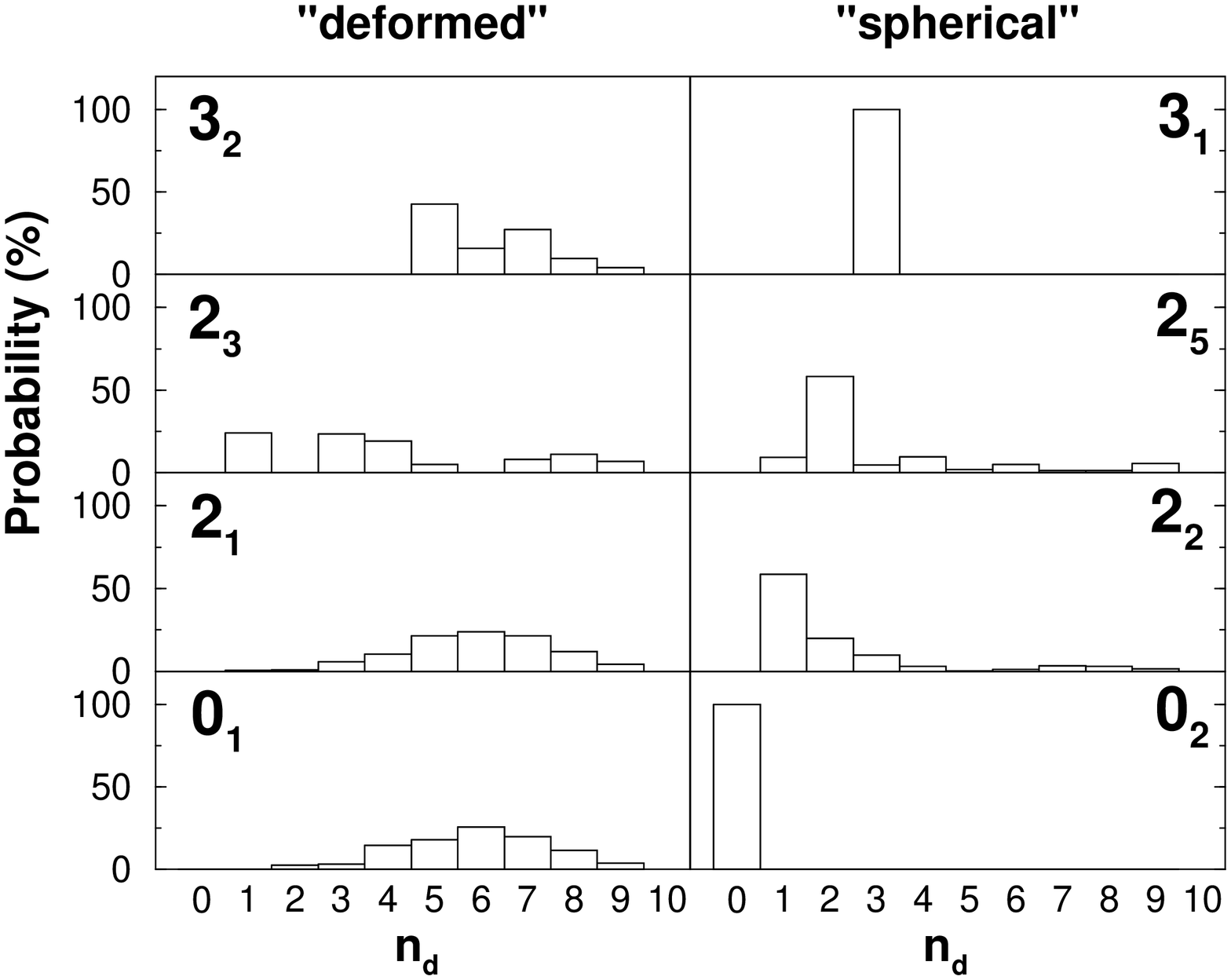}}
\end{center}
\vspace{-1cm}
\caption{Left: spectrum of $H_{int}$, Eq.~(\ref{hint}), 
with $h_2=0.1$, $\beta_0 =1.3$ and $N=10$. 
Right: the number of $d$ bosons ($n_d$) 
probability distribution for selected eigenstates of $H_{int}$.}
\label{fig:2}
\end{figure}
In this case, the intrinsic part 
($H_{int}$) has the equilibrium condensate 
$\vert \,\beta=\beta_0,\gamma=0 ; N\rangle$, 
Eq.~(\ref{cond}), 
as an exact zero-energy eigenstate and has 
an energy surface as in 
Eq.~(\ref{ecri1st}) (with $E_0=0$). 
$H_{int}$ has the form~\cite{lev06}
\ba
H_{int} &=& h_{2}\, 
P^{\dagger}_{2}(\beta_0)\cdot\tilde{P}_{2}(\beta_0)\;\;\; ,\;\;\; 
P^{\dagger}_{2\mu}(\beta_0)= 
\beta_{0}\,s^{\dagger}d^{\dagger}_{\mu} + 
\sqrt{7/2}\,\left( d^{\dagger} d^{\dagger}\right )^{(2)}_{\mu} ~,
\label{hint}
\ea
with
$\tilde{P}_{2\mu}(\beta_0)=(-1)^{\mu}P_{2,-\mu}(\beta_0)$ and 
$h_2=c$. 
By construction $H_{int}$ is rotational-scalar and has the $L$-projected 
states $\vert \beta=\beta_0;N,L\rangle$ of Eq.~(\ref{wfqpt1}) 
as solvable deformed eigenstates with energy $E=0$. 
It has also solvable spherical eigenstates:  
$\vert N,n_d=\tau=L=0 \rangle\equiv \vert s^N\rangle$ and 
$\vert N,n_d=\tau=L=3 \rangle$ with energy $E=0$ 
and $E = 3 h_2\left [\beta_{0}^2 (N-3) + 5 \right ]$ respectively.
Accordingly, the exact spectrum of $H_{int}$ shown in Fig.~2, 
displays a zero-energy deformed ($K=0$) ground band, degenerate with a 
spherical $(n_d=0)$ ground state. The remaining states are either 
predominantly spherical, or deformed states arranged in several excited 
$K=0$ bands below the $\gamma$ band. The coexistence of spherical and 
deformed states is evident in the right portion of Fig.~2, which shows 
the $n_d$ decomposition of wave functions of selected eigenstates of 
$H_{int}$. The ``deformed'' states show a broad $n_d$ distribution typical 
of a deformed rotor structure. The ``spherical'' states show the 
characteristic dominance of single $n_d$ components that one would expect 
for a spherical vibrator. 
\begin{table}[t]
\centering
\caption{
Excitation energies (in units of $E(2^{+}_1)=1$) and B(E2) values 
[in units of $B(E2; 2^{+}_{1}\to 0^{+}_1=1)$] 
for first-order critical-point Hamiltonians 
[high barrier: Eqs.~(\ref{resol}),(\ref{hcol}),(\ref{hint}),
low barrier: Eq.~(\ref{hcriu5su3})], with $N=10$. 
The entries in square brackets [$\ldots$] are estimates based on the 
$L$-projected states, Eq.~(\ref{wfqpt1}), 
with values of $\beta$ 
determined by the global minimum of 
the respective lowest eigenvalue of the potential matrix, 
Eqs.~(\ref{Kij}). 
The rigid-rotor and X(5)~{\protect\cite{iac01}} 
predictions are shown for comparison. 
Adapted from~{\protect\cite{lev06,lev05}}.}
\label{tab1}
\newcommand{\m}{\hphantom{$-$}}
\vspace{2pt}
\begin{tabular}{lcccccc}
\noalign{\smallskip}\hline\noalign{\smallskip}
            &\multicolumn{3}{c}{
high-barrier ($\beta_0=1.3$, $h=0.1$)}
& 
low-barrier
 & rotor & X(5) \\
\noalign{\smallskip}
                & $c_3/h_2=0.05$ & $c_5/h_2 = 0.1$ & $c_6/h_2=0.05$ & 
$\epsilon/\kappa =$ 38.25 &       &  \\
\noalign{\smallskip}\hline\noalign{\smallskip}\noalign{\smallskip}
$E(4^{+}_{1})$  & 
           3.32  [3.32]  & 3.28 [3.28]   & 2.81 [2.87]    & 2.43 [2.46] 
& 3.33  & 2.91 \\
$E(6^{+}_{1})$  & 
           6.98  [6.97]  & 6.74 [6.76]   & 5.43 [5.63]    & 4.29 [4.33]    
& 7.00  & 5.45 \\
$E(8^{+}_{1})$  & 
          11.95 [11.95]  & 11.23 [11.29] & 8.66 [9.04]    & 6.53 [6.56]    
& 12.00 & 8.51 \\
$E(10^{+}_{1})$ & 
          18.26 [18.26]  & 16.58 [16.69] & 12.23 [12.83]  & 9.12 [9.13]  
& 18.33 & 12.07 \\
$E(0^{+}_{2})$  & 
           6.31  [6.30]  & 6.01 [5.93]   & 4.56 [5.03]    & 2.64 [3.30]    
&       & 5.67 \\
\noalign{\smallskip}
\noalign{\smallskip}
$4^{+}_{1}\to 2^{+}_{1}$ &
           1.40 [1.40]   & 1.40 [1.40]  & 1.46 [1.45]     & 1.61 [1.60]  
& 1.43  & 1.58 \\
$6^{+}_{1}\to 4^{+}_{1}$ & 
           1.48 [1.48]   & 1.48 [1.48]  & 1.55 [1.53]     & 1.85 [1.80]   
& 1.57  & 1.98 \\
$8^{+}_{1}\to 6^{+}_{1}$ & 
           1.45 [1.45]   & 1.45 [1.45]   & 1.53 [1.51]    & 1.92 [1.87]
& 1.65  & 2.27 \\
$10^{+}_{1}\to 8^{+}_{1}$  & 
           1.37 [1.37]   & 1.37 [1.37]  & 1.44 [1.42]     & 1.87 [1.86]  
& 1.69  & 2.61 \\
$0^{+}_{2}\to 2^{+}_{1}$ & 
           0.003 [0.003] & 0.003 [0.004] & 0.24 [0.18]    & 0.78 [0.61]  
&       & 0.63 \\
\noalign{\smallskip}\hline
\end{tabular}
\end{table}
Table~\ref{tab1} shows the 
effect of different rotational terms in $H_c$, Eq.~(\ref{hcol}). 
For a high-barrier ($\beta_0=1.3$, $h=0.1$), 
the calculated spectrum resembles a rigid-rotor $(E\sim a_{N}L(L+1)$) for 
the $c_3$-term, a rotor with centrifugal stretching 
$(E\sim a_{N}L(L+1) - b_{N}[L(L+1)]^2)$ for the $c_5$-term, and a 
X(5)-like spectrum for the $c_6$-term. In all cases the B(E2) values are 
close to the rigid-rotor Alaga values. 
Insight of the underlying structure at the critical-point can be gained 
by examining  
the $L$-projected energy surface, obtained by 
the matrix element of the critical Hamiltonian (\ref{resol}) 
in the states~(\ref{wfqpt1}),  
$E^{(N)}_{L}(\beta) = \langle\beta;N,L\vert H_{cri}\vert\beta;N,L\rangle
=\tilde{E}_{L}^{(N)}(\beta)+E_0$, 
\ba
\tilde{E}_{L}^{(N)}(\beta) &=& 
h_2\,(\beta-\beta_0)^2\,\Sigma_{2,L}^{(N)} + 
c_3\left [ L(L+1) - 6D_{1,L}^{(N)}\right ]
+\, c_5\left [ D_{2,L}^{(N)} - \beta^4\,S_{2,L}^{(N)}\right ]
\nonumber\\
&&
+\, c_6\left [ N(N-1) -(1+\beta^2)^2\,S_{2,L}^{(N)}\right ] ~.
\label{eneL}
\ea
Here $D_{1,L}^{(N)}$, $S_{2,L}^{(N)}$, $D_{2,L}^{(N)}$ and 
$\Sigma_{2,L}^{(N)}$ denote the expectation values in 
the states $\vert \beta;N,L\rangle$ 
of $\hat{n}_d$, $\hat{n}_s(\hat{n}_s-1)$, $\hat{n}_d(\hat{n}_d-1)$ 
and $\hat{n}_s\hat{n}_d$ 
respectively. 
The mixing between the coexisting spherical and deformed $L=0$ states, 
$\vert\phi_1\rangle\equiv \vert s^N\rangle$ and 
$\vert\phi_2\rangle\equiv \vert \beta;N,L=0\rangle$, 
can be studied by 
transforming to an orthonormal basis 
\ba
\vert \Psi_1\rangle &=& \vert \phi_1\rangle 
\;\; , \;
\vert \Psi_2\rangle = 
(1-r_{12}^2)^{-1/2}\,
\Bigl (\vert\phi_2\rangle  - r_{12}\,\vert \phi_1 \rangle\Bigr ) ~,
\nonumber\\
r_{12} &=& \langle \phi_1 \vert \phi_2\rangle ~, 
\ea
and examining the $2\times 2$ potential energy matrix, 
$K_{ij}(\beta) = 
\langle \Psi_{i}\vert\, H_{cri}\,\vert \Psi_{j} \rangle$, 
which reads 
\ba
K_{11}(\beta) &=& E_0\;\; , \;\;
K_{12}(\beta) = -c_6\,\beta^2 N(N-1)(1-r_{12}^2)^{-1/2}r_{12} ~,
\nonumber\\
K_{22}(\beta) &=& 
\left [\tilde{E}^{(N)}_{L=0}(\beta) 
+ 2c_6\,\beta^2 N(N-1)\, r_{12}^2\right ]
(1-r_{12}^2)^{-1} +E_0 ~.
\label{Kij}
\ea 
The derived eigenvalues of the matrix 
serve as eigenpotentials, $E^{(\pm)}_{L=0}(\beta)$, and the corresponding 
eigenvectors, $\vert \Phi^{(\pm)}_{L=0}\rangle$, 
are identified with the ground ($0^{+}_1$) and excited ($0^{+}_{i}$) 
$L=0$ states. 
The deformed states $\vert \beta;N,L\rangle$ of Eq.~(\ref{wfqpt1}) 
with $L>0$ are identified 
with excited members of the ground-band ($L^{+}_1$) with energies given by 
$E_{L}^{(N)}(\beta)$, Eq.~(\ref{eneL}).
B(E2) values 
between these states can also be evaluated in 
closed form~\cite{lev06,lev05}. 
The parameter $\beta$ in the indicated wave functions 
and matrix elements is chosen at the global minimum of the lowest 
eigenvalue, $E^{(-)}_{L=0}(\beta)$, of the matrix 
$K_{ij}(\beta)$~(\ref{Kij}). 
As is evident from Table~\ref{tab1}, 
this procedure leads to accurate finite-N estimates of observables 
at the critical-point. 
The characteristic spectra, discussed above, 
of the rotational terms in $H_c$ (\ref{hcol})
can now be understood from 
the $L$-dependence of their respective contributions to 
$E^{(N)}_{L}(\beta)$, Eq.~(\ref{eneL}), and 
$K_{ij}(\beta)$, Eq.~(\ref{Kij}). 
All rotational terms contribute diagonal $L(L+1)$-type 
splitting.
The $c_6$-term 
controls the mixing which is essential for 
obtaining an $X(5)$-like spectrum. 

The same type of analysis can be done~\cite{lev05} for a first-order 
critical-point with a low barrier. A representative Hamiltonian for 
this class is the critical $U(5)$-$SU(3)$ Hamiltonian 
\ba
H_{cri} &=& \epsilon\,\hat{n}_d -\kappa\, Q\cdot Q
\quad ,\quad
\epsilon = 
9\kappa (2N-3)/4 ~.
\label{hcriu5su3}
\ea
Its intrinsic and collective resolution corresponds to the choice 
$h_2=4\kappa$, $\beta_0=1/2\sqrt{2}$, 
$c_3=15\kappa/8$, $c_5=-9\kappa/2$, $c_6=\kappa$, $E_0=-5\kappa N$ 
in Eqs.~(\ref{hcol}),(\ref{hint}). The resulting barrier
is extremely low, $h\approx 10^{-3}$.   
The spectrum and E2 rates, shown in Table~\ref{tab1}, resemble 
the X(5) predictions albeit finite-N modifications. 

\section{SECOND-ORDER CRITICAL-POINT}

The energy surface for a second-order transition between spherical 
and $\gamma$-unstable deformed shapes is independent of $\gamma$ 
and has the following form at the critical-point 
\ba
E_{cri}(\beta) &=&
E_0 + c\,N(N-1)g(\beta)\;\;\; , \;\;\;
g(\beta) = \beta^4\,(1+\beta^2)^{-2} ~.
\label{ecri2nd}
\ea
As shown in Fig~1(b), 
$E_{cri}(\beta)$ exhibits a flat-bottomed behavior ($\sim \beta^4$) 
for small $\beta$, hence the 
single minimum at $\beta=0$ is not well-localized and experiences 
large $\beta$ fluctuations. 
The relevant critical Hamiltonian (\ref{resol}) has $O(5)$ 
symmetry. Its intrinsic part is given by
\ba
H_{int} = t_0\,\hat{n}_d(\hat{n}_d-1) ~,
\label{hint2nd}
\ea
with $t_0=c$, and the collective part has the form given in 
Eq.~(\ref{hcol}). The states $\vert\beta;N,\tau,L\rangle$ of 
Eq.~(\ref{wfqpt2}) constitute suitable wave functions for yrast states 
at the critical-point, with the value of $\beta$ chosen at 
the global minimum of the $O(5)$-projected energy surface, 
$E^{(N)}_{\tau,L}(\beta) = 
\langle\beta;N,\tau,L\vert H_{cri}\vert\beta;N,\tau,L\rangle$, 
\ba
E_{\tau,L}^{(N)}(\beta) &=& 
t_{0}\,D^{(N)}_{2,\tau} + 
c_3\left [ L(L+1) - 6D_{1,\tau}^{(N)}\right ]
+\, c_5\left [ \tau(\tau+3) -4D_{1,\tau}^{(N)}\right ]
\nonumber\\
&&
+\, c_6\left [ N(N-1) -(1+\beta^2)^2\,S_{2,\tau}^{(N)}\right ] + E_0 ~.
\label{enetau}
\ea
Here $D_{1,\tau}^{(N)}$, $D_{2,\tau}^{(N)}$ and $S_{2,\tau}^{(N)}$ 
denote the expectation values in 
the states $\vert \beta;N,\tau,L\rangle$ 
of $\hat{n}_d$, $\hat{n}_d(\hat{n}_d-1)$ 
and $\hat{n}_s(\hat{n}_s-1)$ respectively. 
This procedure has been tested in \cite{levgin03} for 
a representative Hamiltonian of this class, namely, 
the $U(5)$-$O(6)$ critical 
Hamiltonian, $H_{cri} = \epsilon\,\hat{n}_d + A\hat{P}_{6}$, 
with $\epsilon = (N-1)A$.
Its intrinsic-collective resolution corresponds to the choice
$t_0=A,\, c_3=0,\, c_5=-A/2,\,c_6=A/4,\, E_0=AN(N-1)/4$.
in Eqs.~(\ref{hcol}),(\ref{hint2nd}).


\begin{thebibliography}{9}
\bibitem{bender03} 
M.~Bender, P.~H.~Heenen and P.G.~Reinhard, 
Rev. Mod. Phys. 75 (2003) 121.

\bibitem{iac00}
F.~Iachello, 
Phys. Rev. Lett. 85 (2000) 3580. 

\bibitem{iac01}
F.~Iachello, 
Phys. Rev. Lett. 87 (2001) 052502. 

\bibitem{caszam0001}
R.~F.~Casten and N.~V.~Zamfir, 
Phys. Rev. Lett. 85 (2000) 3584; 87 (2001) 052503.

\bibitem{ibm}
F.~Iachello and A.~Arima,
The Interacting Boson Model, 
Cambridge Univ. Press, Cambridge 1987.

\bibitem{gino80}
J.~N.~Ginocchio and M.~W.~Kirson, 
Phys. Rev. Lett. 44 (1980) 1744. 

\bibitem{diep80}
A.~E.~L.~Dieperink, O.~Scholten and F.~Iachello, 
Phys. Rev. Lett. 44 (1980) 1747.

\bibitem{kirlev85}
M.~W.~Kirson and A.~Leviatan,
Phys. Rev. Lett. 55 (1985) 2846.

\bibitem{lev87}
A.~Leviatan, Ann. Phys. (NY) 179 (1987) 201.

\bibitem{iaczam04}
F.~Iachello and N.~V.~Zamfir, 
Phys. Rev. Lett. 92 (2004) 212501 and references therein.

\bibitem{lev06}
A.~Leviatan, (2006) submitted for publication.

\bibitem{lev05} 
A.~Leviatan, 
Phys. Rev. C 72 (2005) 031305.

\bibitem{levgin03}
A.~Leviatan and J.~N.~Ginocchio, 
Phys. Rev. Lett. 90 (2003) 212501.

\end{thebibliography}
\end{document}